\documentclass[twocolumn,prd,aps,groupedaddress,showpacs,nofootinbib]{revtex4}

\usepackage{amsmath}
\usepackage{amssymb}
\usepackage{latexsym}
\usepackage{graphicx}
\usepackage{hyperref}
\usepackage{mathrsfs}
\usepackage{color}
\usepackage{bm}
\usepackage[caption=false]{subfig}

\begin{document}

\title{Reconstructing the interaction between dark energy and dark matter using Gaussian processes}
\author{Tao Yang$^1$}
\email{yangtao@itp.ac.cn}
\author{Zong-Kuan Guo$^1$}
\email{guozk@itp.ac.cn}
\author{Rong-Gen Cai$^1$}
\email{cairg@itp.ac.cn}

\affiliation{\it $^1$State Key Laboratory of Theoretical Physics, Institute of
Theoretical Physics, Chinese Academy of Sciences, P.O. Box 2735,
Beijing 100190, China }

\pacs{95.36.+x, 98.80.-k, 98.80.Es}

\begin{abstract}
 We present a nonparametric approach to reconstruct the interaction between dark energy and dark matter directly from SNIa Union 2.1 data using Gaussian processes, which is a fully Bayesian approach for smoothing data. In this method, once the equation of state ($w$) of dark energy is specified, the interaction can be reconstructed as a function of redshift. For the decaying vacuum energy case with
 $w=-1$, the reconstructed interaction is consistent with the standard $\Lambda$CDM model, namely, there is no evidence for the interaction. This also holds for the  constant $w$ cases from $-0.9$ to $-1.1$ and for the Chevallier-Polarski-Linder (CPL) parametrization case. If the equation of state deviates obviously from $-1$, the reconstructed interaction exists at $95\%$ confidence level. This shows the degeneracy between the interaction and the equation of state of dark energy when they get constraints from the observational data.
 \end{abstract}
\maketitle
\section{Introduction \label{sec:introduction}}

It has been more than fifteen years since the universe was found in accelerating expansion~\cite{Riess:1998cb,Perlmutter:1998np}.
However, it is fair to say that its origin is still not yet clear. A possible explanation of this cosmic acceleration is provided by the introduction of a fluid with negative pressure called dark energy (DE). The simplest dark energy candidate is the cosmological constant $\Lambda$ with the equation of state $w =-1$. The tiny cosmological constant together with the  cold dark matter (CDM) (called the $\Lambda$CDM model) turned out to be the standard model which fits the current observational data sets consistently. In spite of this success, however, it is faced with the fine-tuning problem~\cite{Weinberg:2000yb} and the coincidence problem. The former arises from the fact that the present-time observed value for the vacuum energy density is more than 120 orders of magnitude smaller than the naive estimate from quantum field theory. The later is the question why we live in such a special moment that the densities of dark energy and dark matter are of the same order.

Many attempts have been made to tackle those issues, including introducing ``dynamical" dark energy or modifying general relativity at the cosmic scales. 
 In particular, to alleviate the coincidence problem, an interaction between dark energy and dark matter has been considered extensively in the literature, for example, in~\cite{Farrar:2003uw,Cai:2004dk,Wang:2006qw,Amendola:2006dg,Guo:2007zk,He:2008tn,CalderaCabral:2008bx,Cai:2009ht,Honorez:2010rr,Bessada:2013maa,Yang:2014gza,Salvatelli:2014zta,Li:2014cee,Skordis:2015yra} and references therein.  However, usually the interaction form has to be assumed. The assumed form will lead to a bias when observational data are used to give the constraint on the interaction. In \cite{Cai:2009ht} the authors investigated the possible interaction in a way independent of specific interacting forms by dividing the whole range of redshift into a few bins and setting the interacting term to be a constant in each redshift bin. It was found that the interaction is likely to cross the noninteracting line and has an oscillation behavior.  Recently,  Salvatelli {\it et al.}~\cite{Salvatelli:2014zta} showed that the null interaction is excluded at $99\%$ confidence level (C.L.) when they added the redshift-space distortions (RSD) data to  the Planck data for the decaying vacuum energy model (a class of interaction of dark sectors). They parametrized the interaction term to be of the form $qH\rho$ and also subdivided the redshift into four bins with $q(z)=q_i (i=1, \ldots, 4)$. More recently, the authors of Ref.~\cite{Wang:2015wga} have reconstructed the temporal evolution of the coupling strength between dark matter and vacuum energy, $\alpha(a)$ in a nonparametric Bayesian approach using the combined observational data sets from the cosmic microwave background radiation, supernovae and large scale structure. It was found that an evolving interaction can remove some of the tensions between different types of data sets, and is favored at $\sim95\%$ C.L. if the baryon acoustic oscillations measurements of the BOSS Lyman-$\alpha$ forest sample are included. Thus, it is quite interesting to see whether there exists some signature of the interaction from the observational data in a model-independent way.

 In this paper we present a nonparametric approach to reconstruct the interaction term between dark energy and dark matter directly from the observational data using Gaussian processes (GP). GP is a model independent method to smooth the data. We set the nonparameterized interaction term $Q(z)$ as a function of redshift and reconstruct it from SNIa Union 2.1 data sets. We will consider three cases, the decaying vacuum energy case with $w=-1$, the $w$CDM model and the Chevallier-Polarski-Linder (CPL) parametrization of dark energy, respectively, and pay special attention to the first case as discussed in \cite{Salvatelli:2014zta}.

This paper is organized as follows. In Sec.~\ref{sec:model}  we give the interacting models of dark energy and dark matter in a flat universe. In Sec.~\ref{sec:methods}, we briefly introduce the Gaussian processes method and create a mock data set to demonstrate the reliability  of the GP reconstruction method. Then we apply it to the real data Union 2.1 in the decaying vacuum energy case, followed by different values of $w$ for comparison and the CPL case. We give some discussions and make conclusions in Sec.~\ref{sec:discussion}.

\section{ The model \label{sec:model}}


In a flat universe with an interaction between dark energy and dark matter, the Friedmann equation describing the evolution of the universe is given by
\begin{equation}
{H^2} = \frac{{8\pi G}}{3}({\rho _m} + {\rho _{DE}}),
\end{equation}
where $\rho_m$ denotes the energy density of dark matter and $\rho_{DE}$ the energy density of dark energy. However, the conservation equations are changed to be
\begin{equation}
\dot\rho_m+3H\rho_m =  - Q,
\end{equation}
\begin{equation}
\dot\rho_{DE}+3H(1+w)\rho_{DE} = Q,
\end{equation}
where $w$ is the equation of state of DE, $H$ is the expansion rate of the universe, and $Q$ describes the interaction between dark matter and dark energy. When $Q=0$ and $w=-1$, it recovers the standard $\Lambda$CDM model. Unlike most of the ways to parametrize the interaction term $Q$ using such a form $qH\rho$~\cite{Cai:2004dk,Salvatelli:2014zta}, here we use $Q(z)$ itself and want to reconstruct it directly from data using a model-independent method.
As assumed in~\cite{Salvatelli:2014zta}, the 4-vector $Q^\nu$ is proportional to the 4-velocity of dark matter.
Hence the perturbation of the interaction is not considered in this paper.

Combining the Friedmann equation and the conservation equations ($8\pi G=1$ throughout the paper), we can obtain
\begin{align}
- wQ &~= 2\left(HH{'^2} + {H^2}H'' - \frac{{w'}}{w}{H^2}H'\right){(1 + z)^2} \nonumber\\
         &~~ - \left[2(5 + 3w){H^2}H' - \frac{{3{H^3}w'}}{w}\right](1 + z) \nonumber\\
         &~~ + 9(1 + w){H^3},
\end{align}
where the prime denotes the derivative with respect to redshift $z$. Note that here we also assume $w$ is time dependent. For convenience, we use a dimensionless $q$ to characterize the interaction, i.e., $Q= q H_0^3$. Note that here $q$ is not the usual deceleration.  In this case, we have
\begin{align}
- wq{H_0}^3 &~= 2\left(HH{'^2} + {H^2}H'' - \frac{{w'}}{w}{H^2}H'\right){(1 + z)^2} \nonumber\\
             &~~ - \left[2(5 + 3w){H^2}H' - \frac{{3{H^3}w'}}{w}\right](1 + z) \nonumber\\
             &~~ + 9(1 + w){H^3}.
\label{equa:qH}
\end{align}
The  luminosity distances at redshift $z$ can be expressed as
\begin{equation}
{d_L}(z) = \frac{{c(1 + z)}}{{{H_0}}}\int_0^z {d{z^*}\frac{{{H_0}}}{{H({z^*})}}}.
\label{equa:dl}
\end{equation}
 Writing $D(z)=(H_0/c)(1+z)^{-1} d_L(z)$ as the normalized comoving distance, we can arrive at
\begin{align}
- wq = &~2\left(\frac{{3D'{'^2}}}{{D{'^5}}} - \frac{{D'''}}{{D{'^4}}} + \frac{{w'D''}}{{wD{'^4}}}\right){(1 + z)^2} \nonumber\\
         &~~ + \left[2(5 + 3w)\frac{{D''}}{{D{'^4}}} + \frac{{3w'}}{{wD{'^3}}}\right](1 + z) \nonumber\\
         &~~ + \frac{{9(1 + w)}}{{D{'^3}}}.
\label{equa:qD}
\end{align}
From this, we see that using the observed distance-redshift relationship $D(z)$, one can reconstruct the interaction, once the equation of state $w$ of dark energy is given.

\section{reconstruction method
 \label{sec:methods}}
In order to reconstruct the interaction using current data sets, we should find a model-independent method to reconstruct $D(z)$ and its derivatives. While there are several methods such as principle component analysis~\cite{Huterer:2002hy,Shapiro:2005nz,Clarkson:2010bm}, Gaussian smoothing~\cite{Shafieloo:2005nd,Shafieloo:2007cs} and Gaussian processes~\cite{Holsclaw:2010nb,Holsclaw:2010sk,Holsclaw:2011wi,Seikel:2012uu},  in this paper we will reconstruct $D(z)$ and its derivatives more precisely by using the GP method.

\subsection{Gaussian processes}

The Gaussian processes allows one to reconstruct a function from data without assuming a parametrization for it. We use Gaussian processes in Python (GaPP)~\cite{Seikel:2012uu} to derive our GP results. This GP code has been applied in many papers~\cite{Seikel:2012uu,Seikel:2012cs,Seikel:2013fda,Yahya:2013xma,Busti:2014dua,Zhang:2014eux,Busti:2015aqa}. The distribution over functions  provided by GP is suitable to describe the observed data. At each point $z$, the reconstructed function $f(z)$ is also a Gaussian distribution with a mean value and Gaussian error. The functions at different points $z$ and $\tilde{z}$ are related by a covariance function $k(z,\tilde{z})$, which only depends on a set of hyperparameters $\ell$ and $\sigma_f$. Here $\ell$ gives a measure of the coherence length of the correlation in $x$-direction and $\sigma_f$ denotes the overall amplitude of the correlation in the $y$-direction.
Both of them will be optimized by GP with the observed data set.
In contrast to actual parameters, GP does not specify the form of the reconstructed function. Instead it characterizes the typical changes of the function.

The different choices of the covariance function may affect the reconstruction to some extent. The covariance function is usually adopted as the squared exponential form~\cite{Seikel:2012uu}
\begin{equation}
k(z,\tilde{z})={\sigma_f}^2 \exp\Big(-\frac{(z-\tilde{z})}{2\ell^2}\Big).
\end{equation}
But it is not always the suitable choice. Here we take the Mat\'{e}rn ($\nu = 9/2$) covariance function
\begin{align}
k(z,\tilde z) = &~{\sigma _f}^2\exp\left( - \frac{{3\left| {z - \tilde z} \right|}}{\ell }\right) \nonumber \\
      &~~\times\Big[1 + \frac{{3\left| {z - \tilde z} \right|}}{\ell } + \frac{{27{{(z - \tilde z)}^2}}}{{7{\ell ^2}}} \nonumber \\
     &~~ + \frac{{18{{\left| {z - \tilde z} \right|}^3}}}{{7{\ell ^3}}} + \frac{{27{{(z - \tilde z)}^4}}}{{35{\ell ^4}}}\Big],
\end{align}
according to the analysis made in~\cite{Seikel:2013fda}, where they consider various assumed models and many realizations of mock data sets for a test and conclude that the Mat\'{e}rn ($\nu=9/2$) covariance function can lead to more reliable results than all others when applying GP to reconstructions using $D$ measurements.

Following Refs.~\cite{Seikel:2012uu,Yahya:2013xma}, in which the detailed technical description of GP can be found, we reconstruct the interaction between dark energy and dark matter using the SNIa Union 2.1 data set~\cite{Suzuki:2011hu}. Before that we will first show the reliability of the GP method.

\begin{figure*}
\includegraphics[width=0.35\textwidth]{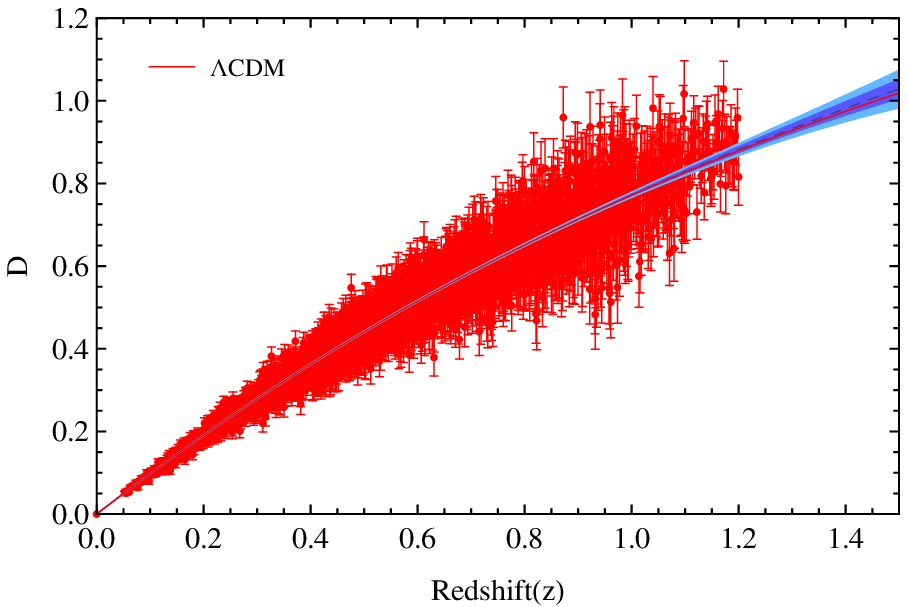}\quad
\includegraphics[width=0.35\textwidth]{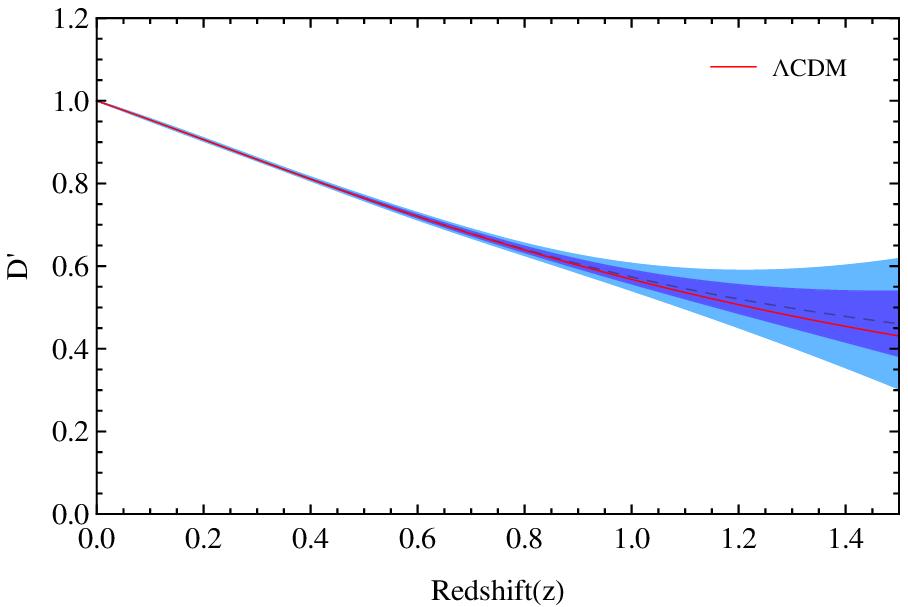}\\
\includegraphics[width=0.35\textwidth]{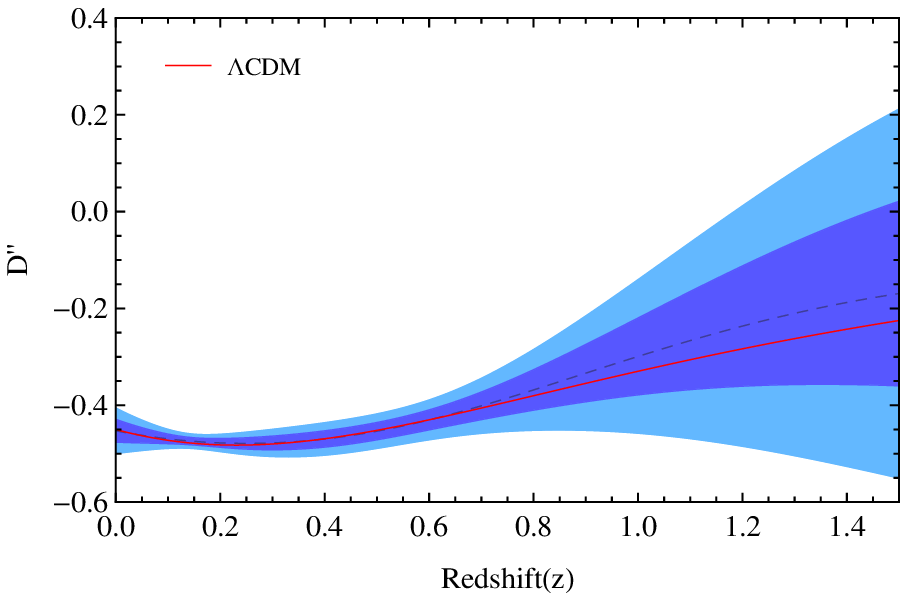}\quad
\includegraphics[width=0.35\textwidth]{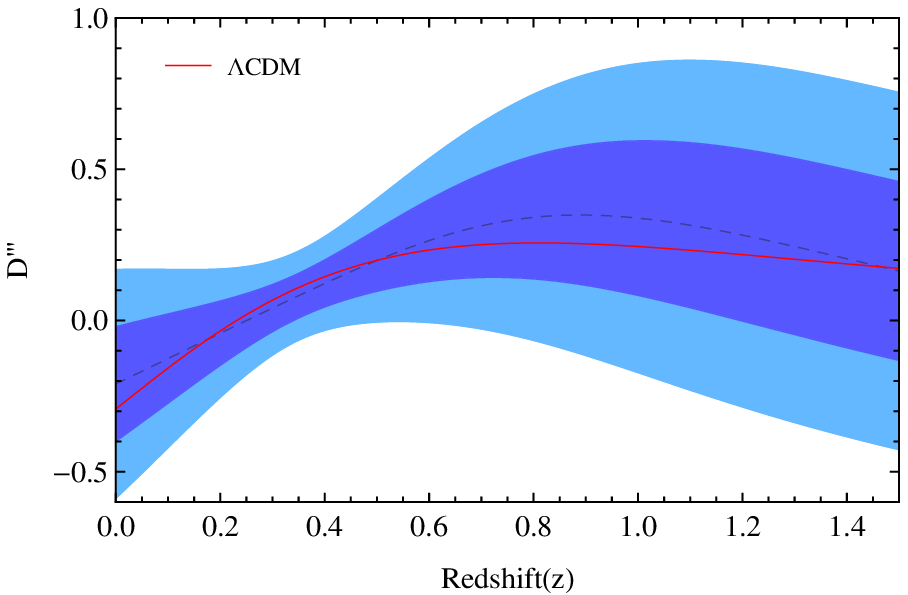}
\caption{\label{fig:mockDLCDM}Gaussian processes reconstruction of $D(z)$, $D'(z)$ ({\em top}), and $D''(z)$, $D'''(z)$ ({\em bottom}) obtained from a mock data set of future DES and assuming the $\Lambda$CDM model with $\Omega_{m0}=0.3$ (red line). The dashed blue line is the mean of the reconstruction and the shaded blue regions are the $68\%$ and $95\%$ C.L. of the reconstruction, respectively.}
\end{figure*}

\begin{figure*}
\includegraphics[width=0.6\textwidth]{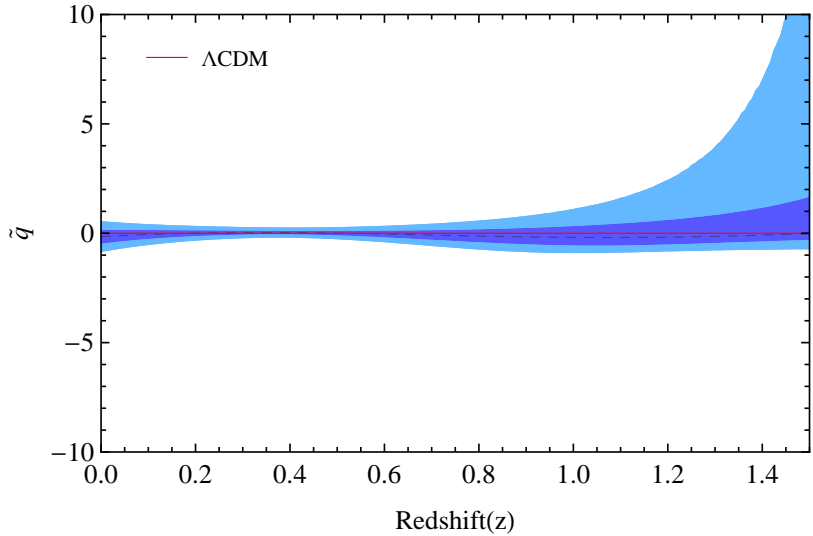}
\caption{\label{fig:mockqLCDM}Reconstruction of $\tilde{q}(z)$ (dashed line) from the mock data set of future DES and assuming the $\Lambda$CDM model with $\Omega_{m0}=0.3$. The shaded blue regions are the $68\%$ and $95\%$ C.L. of the reconstruction.}
\end{figure*}

\begin{figure*}
\includegraphics[width=0.35\textwidth]{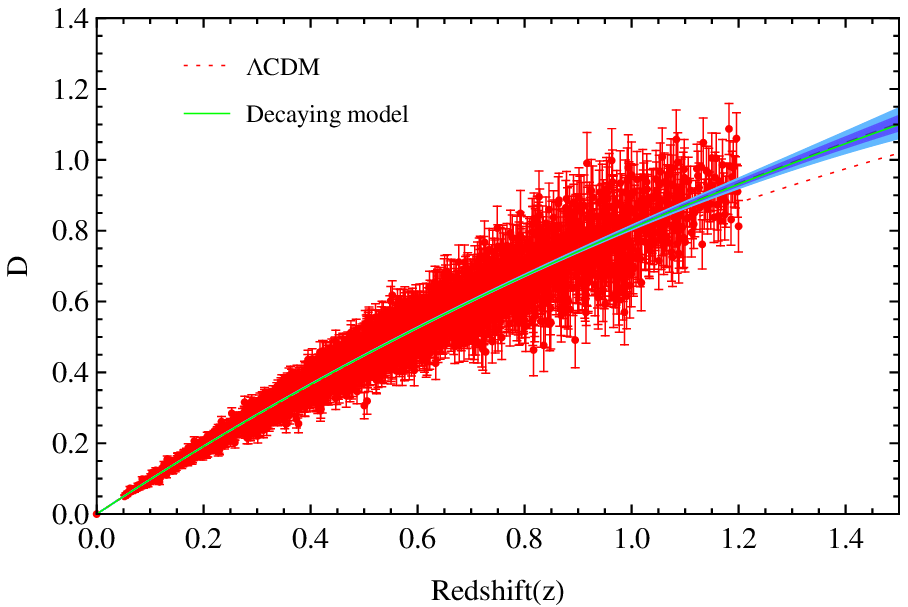}\quad
\includegraphics[width=0.35\textwidth]{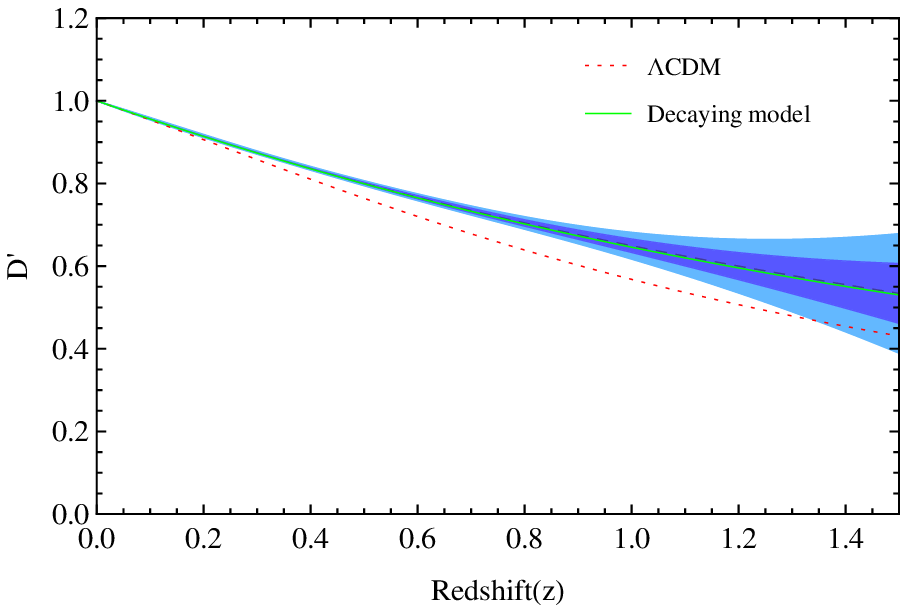}\\
\includegraphics[width=0.35\textwidth]{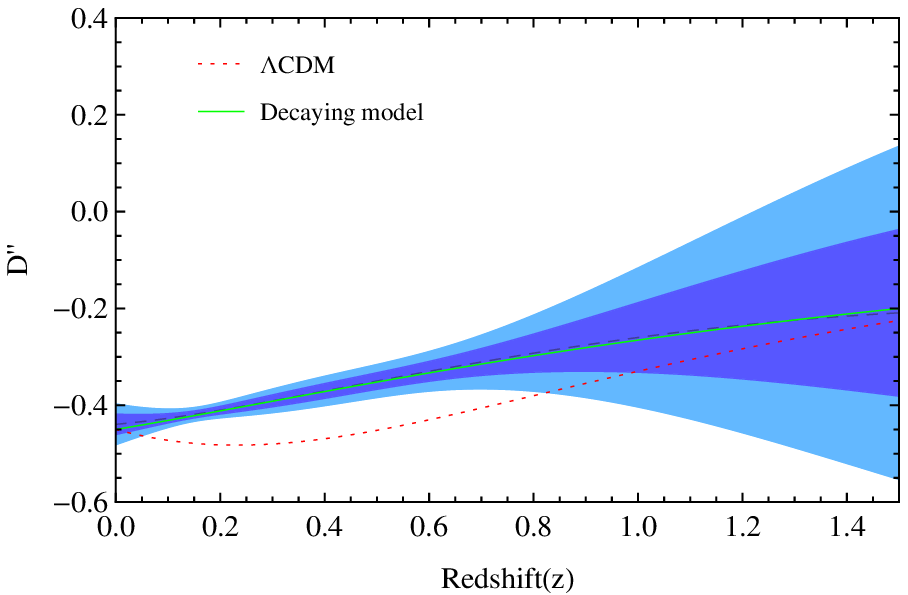}\quad
\includegraphics[width=0.35\textwidth]{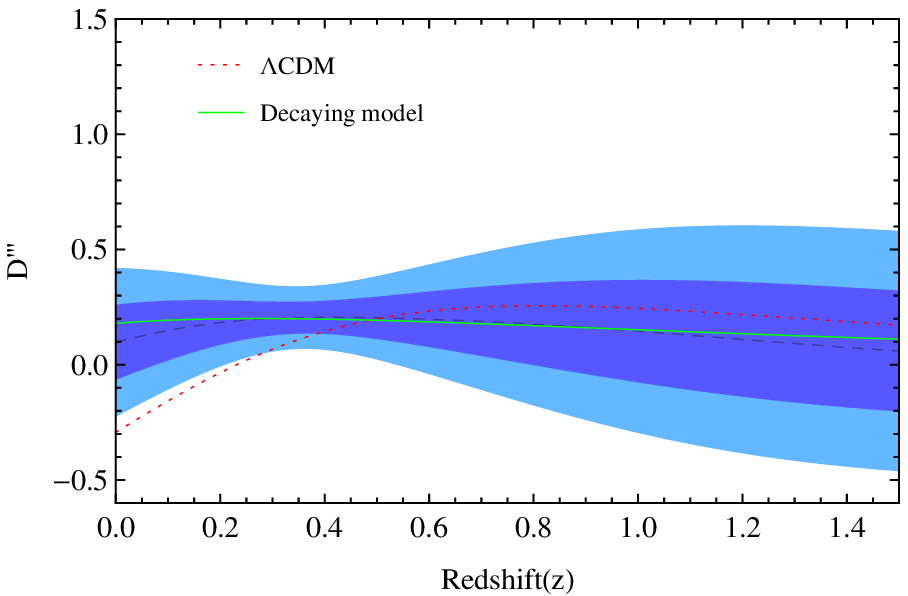}
\caption{\label{fig:mockDdecay}Gaussian processes reconstruction of $D(z)$, $D'(z)$ ({\em top}), and $D''(z)$, $D'''(z)$ ({\em bottom}) obtained from a mock data set of future DES and assuming a toy decaying vacuum model: $\rho_{DE}=3\alpha H$ with $w=-1$ and $\Omega_{m0}=0.3$ (green line). The dashed blue line is the mean of the reconstructions and the shaded blue regions are the $68\%$ and $95\%$ C.L. of the reconstruction, respectively. The $\Lambda$CDM model is also shown (red dotted).}
\end{figure*}

\begin{figure*}
\includegraphics[width=0.6\textwidth]{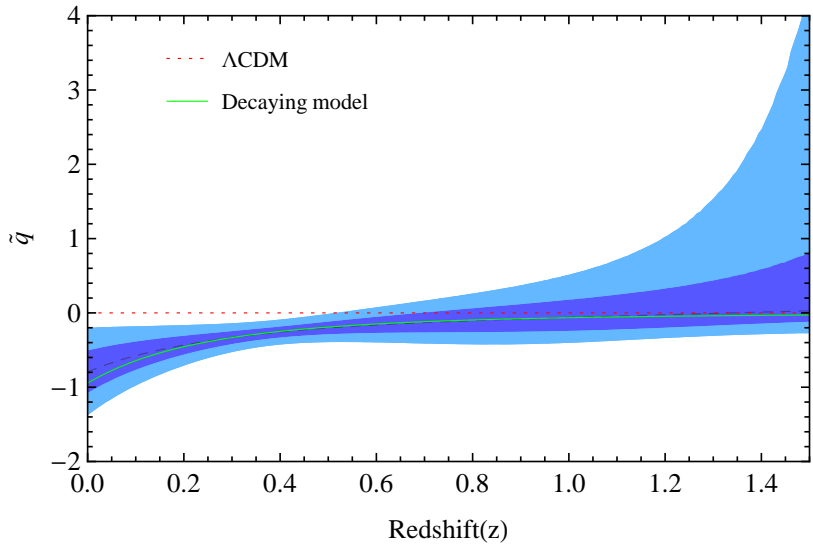}
\caption{\label{fig:mockqdecay}Reconstruction of $\tilde{q}(z)$ from a mock data set of future DES and assuming a decaying-vacuum model: $\rho_{DE}=3\alpha H$ with $w=-1$ and $\Omega_{m0}=0.3$ (green line). The shaded blue regions are the $68\%$ and $95\%$ C.L. of the reconstruction. The $\Lambda$CDM model is also shown (red dotted).}
\end{figure*}

\begin{figure*}
\includegraphics[width=0.35\textwidth]{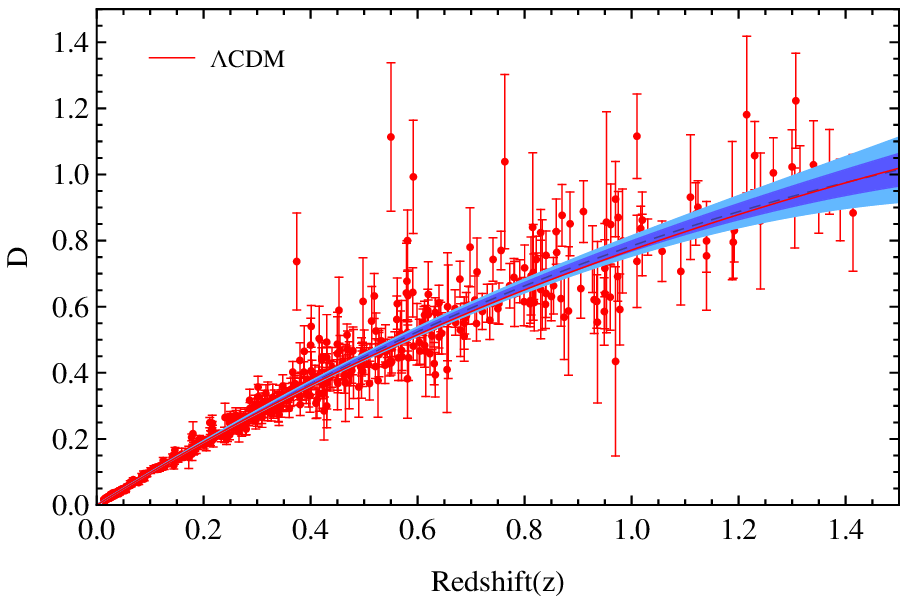}\quad
\includegraphics[width=0.35\textwidth]{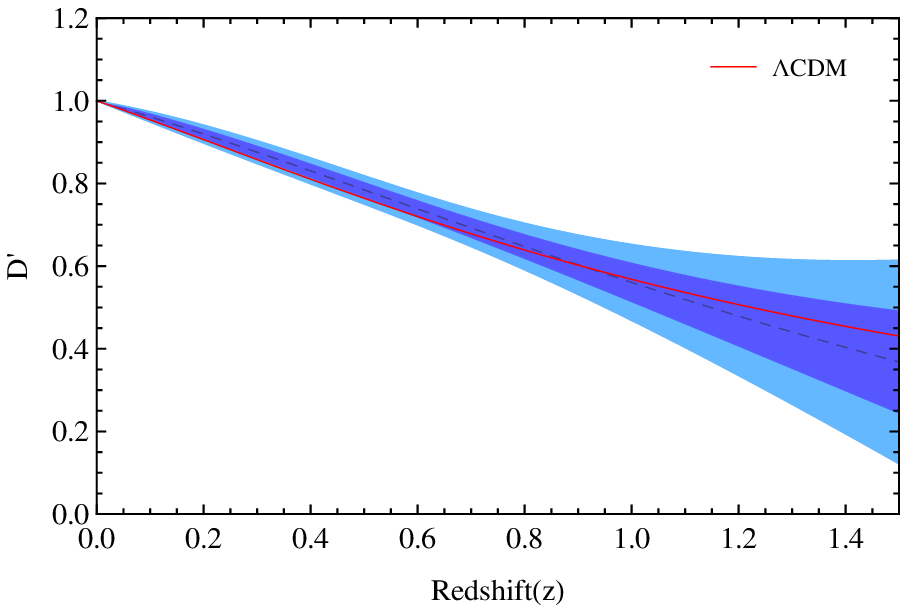}\\
\includegraphics[width=0.35\textwidth]{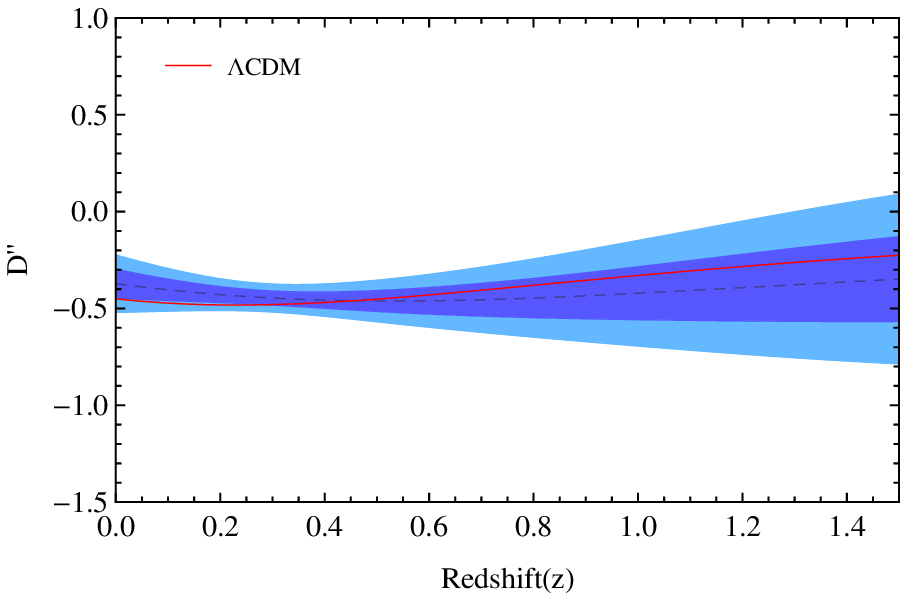}\quad
\includegraphics[width=0.35\textwidth]{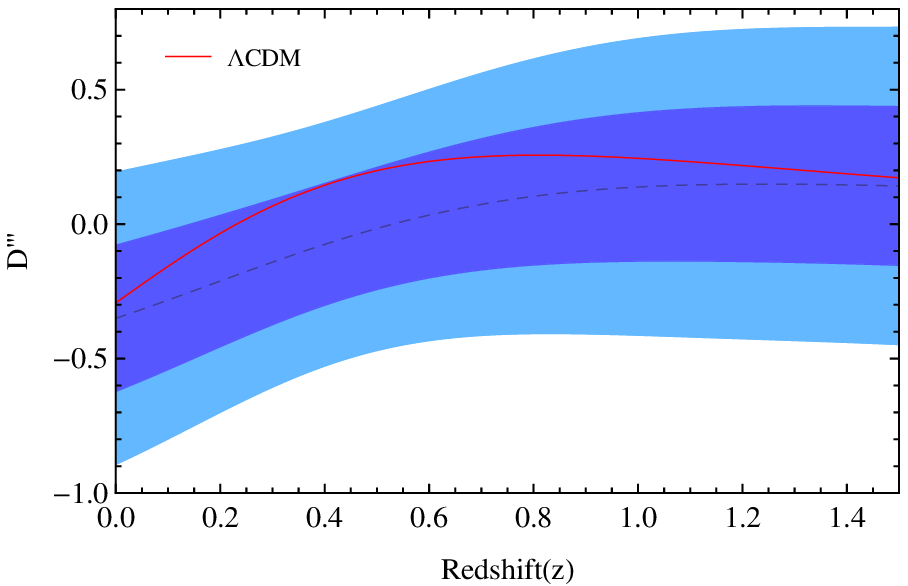}
\caption{\label{fig:Dunion2.1}Gaussian precesses reconstruction of $D(z)$, $D'(z)$ ({\em top}), and $D''(z)$, $D'''(z)$ ({\em bottom}) obtained from Union 2.1 data sets. The shaded blue regions are the $68\%$ and $95\%$ C.L. of the reconstruction. The $\Lambda$CDM model (red line) is also shown.}
\end{figure*}

\begin{figure*}
\includegraphics[width=0.6\textwidth]{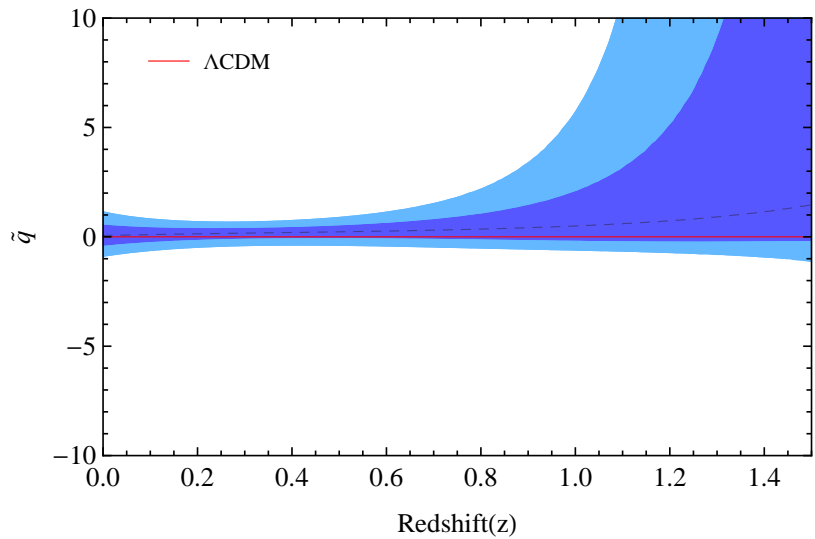}
\caption{\label{fig:qunion2.1}Reconstruction of $\tilde{q}(z)$ from Union 2.1 data sets. The shaded blue regions are the $68\%$ and $95\%$ C.L. of the reconstruction. The red line corresponds to the $\Lambda$CDM model.}
\end{figure*}

\begin{figure*}
\subfloat[$w=-0.7$]{
\includegraphics[width=0.4\textwidth]{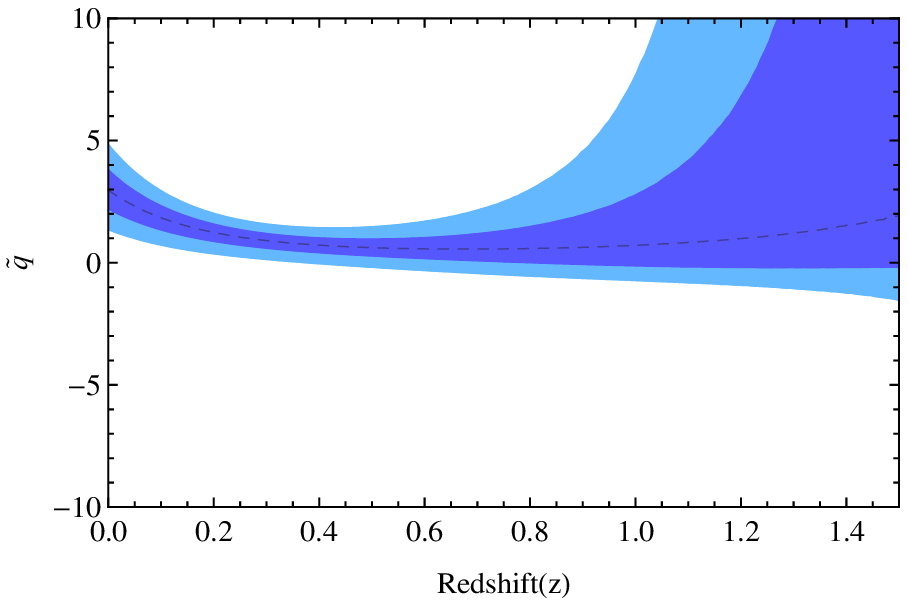}}\quad
\subfloat[$w=-0.8$]{
\includegraphics[width=0.4\textwidth]{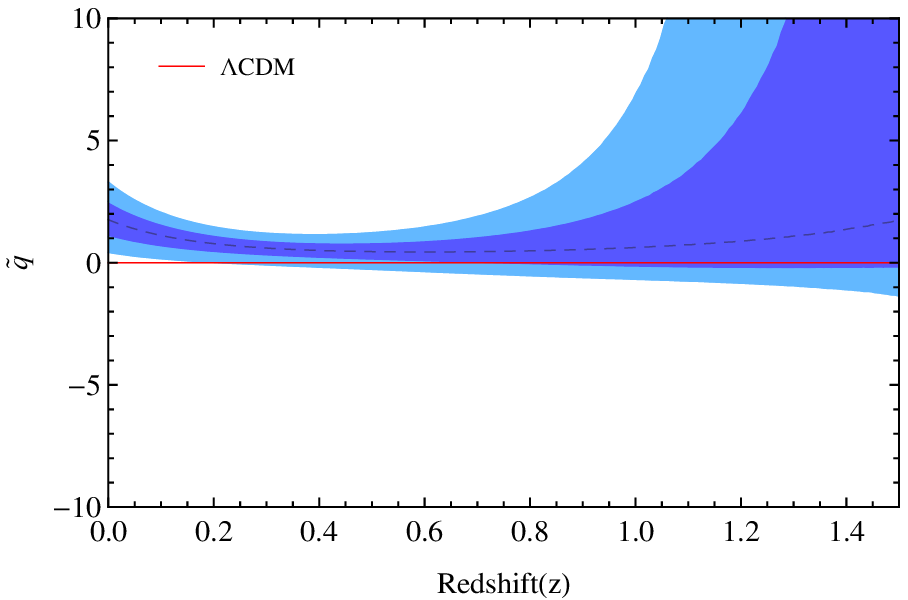}}\\
\subfloat[$w=-0.9$]{
\includegraphics[width=0.4\textwidth]{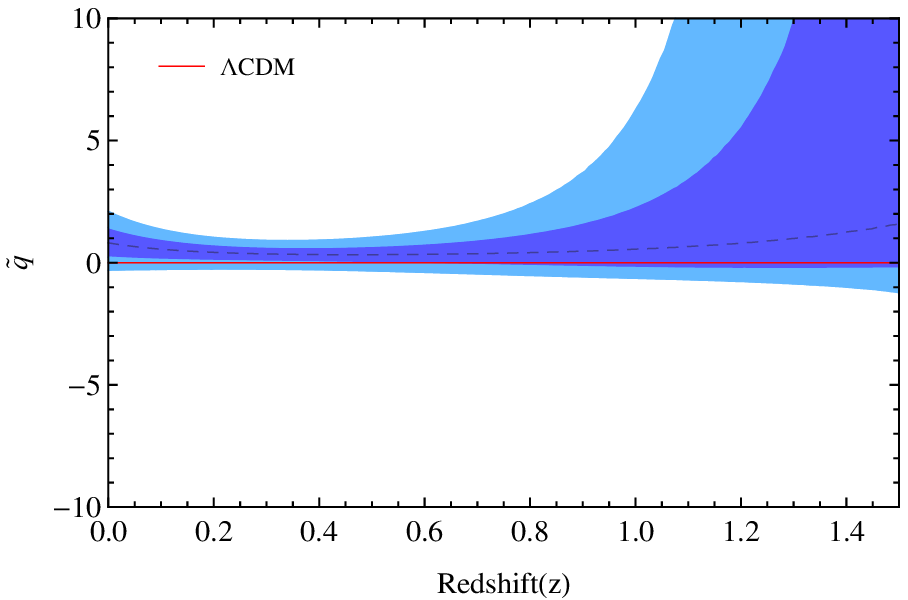}}\quad
\subfloat[$w=-1.006\pm0.045$]{
\includegraphics[width=0.4\textwidth]{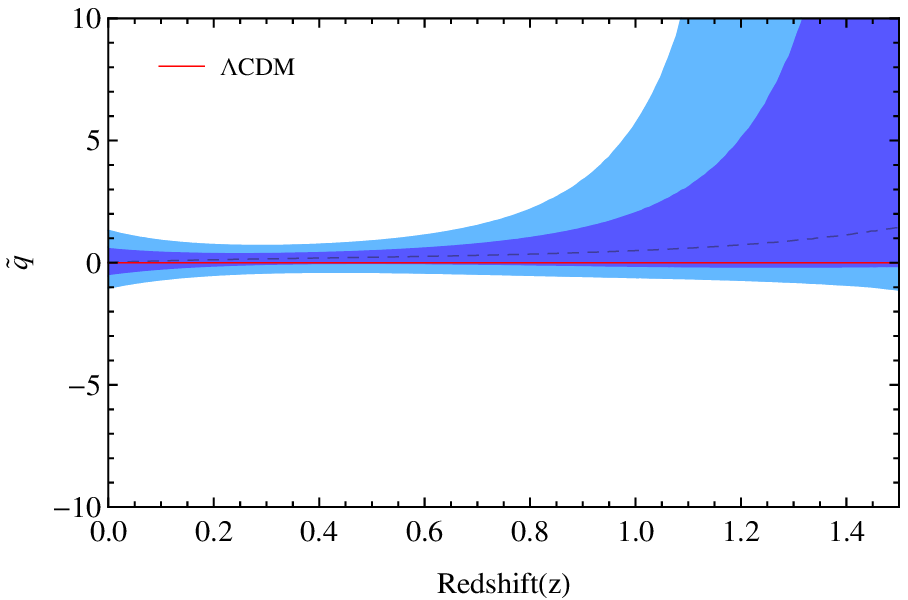}}\\
\subfloat[$w(a)=w_0+w_a(1-a)$]{
\includegraphics[width=0.4\textwidth]{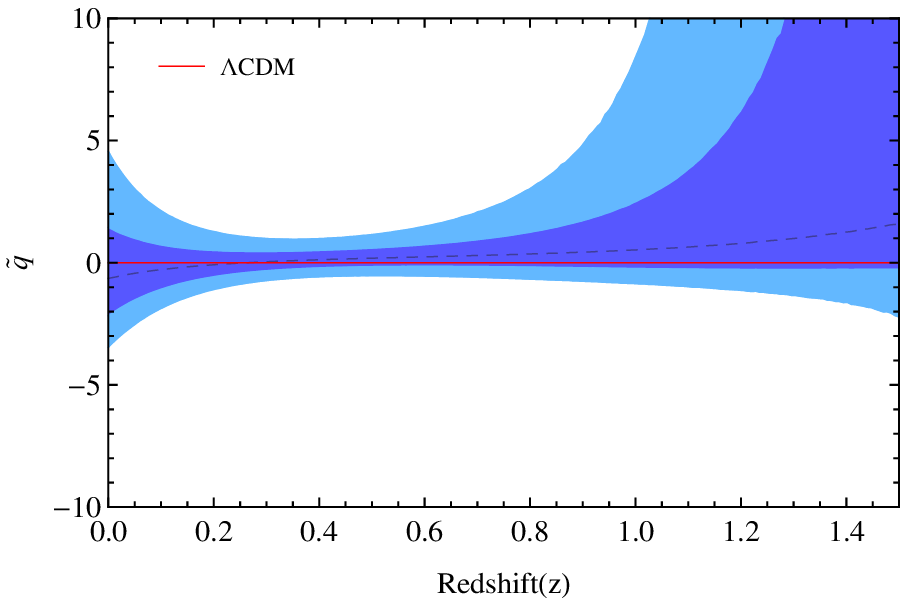}}\quad
\subfloat[$w=-1.1$]{
\includegraphics[width=0.4\textwidth]{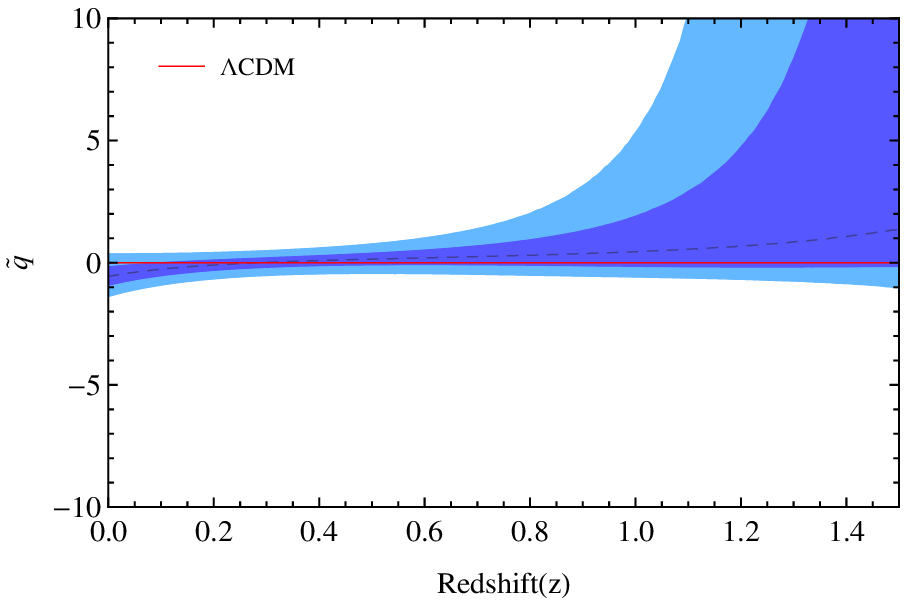}}\\
\subfloat[$w=-1.2$]{
\includegraphics[width=0.4\textwidth]{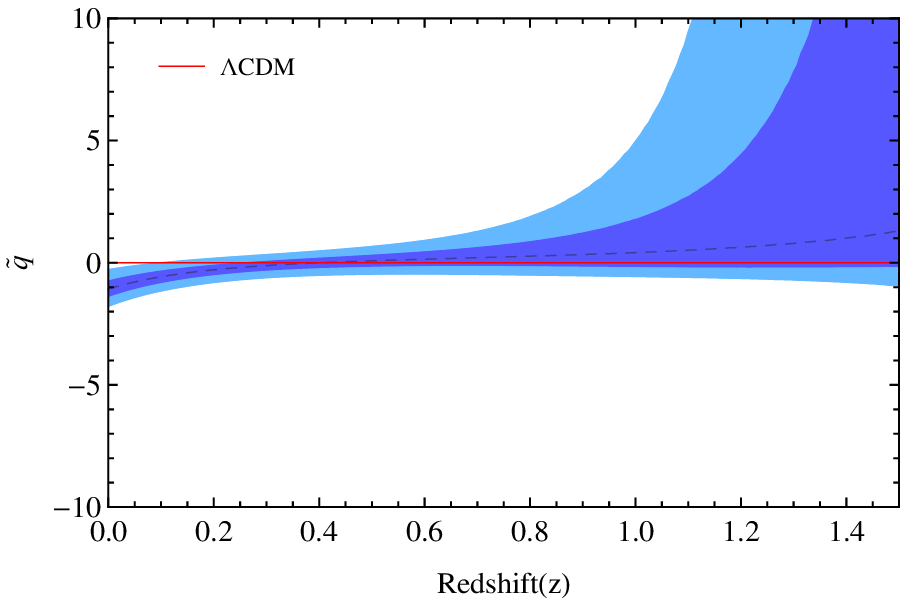}}\quad
\subfloat[$w=-1.3$]{
\includegraphics[width=0.4\textwidth]{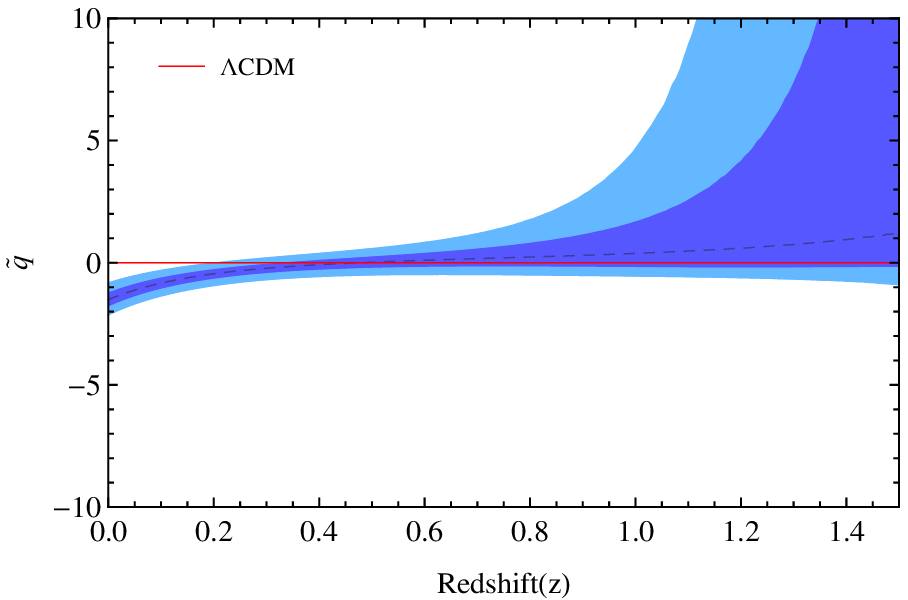}}
\caption{\label{fig:qofotherw}Reconstructions of $\tilde{q}(z)$ for the $w$CDM model and CPL parametrization. (a) $w=-0.7$; (b) $w=-0.8$; (c) $w=-0.9$; (d) $w=-1.006\pm0.045$ from Planck 2015; (e) $w(a)=w_0+w_a(1-a)$ with $w_0=-1.046_{-0.170}^{+0.179}$ and $w_a=0.14_{-0.76}^{+0.60}$; (f) $w=-1.1$; (g) $w=-1.2$; (h) $w=-1.3$; The shaded blue regions are the $68\%$ and $95\%$ C.L. for the reconstruction.}
\label{errorfig}
\end{figure*}

\subsection{Mock data}

To demonstrate the ability of the GP method to distinguish different models and recover the correct behaviors of the models, we create mock data sets of future SNIa according to the Dark Energy Survey (DES)~\cite{Bernstein:2011zf} for two fiducial models: the standard $\Lambda$CDM model and a toy decaying vacuum model: $\rho_{DE}=3\alpha H$ with $w=-1$. Here we set $\Omega_{m0}=0.3$ for both.

For the $\Lambda$CDM model, it is straightforward to calculate the Hubble parameter $H/H_0$ and then obtain the simulated data of $D(z)$ easily. Because there is no interaction, we just want to check whether the reconstruction with this simulated data can recover $q=0$.

As for the decaying vacuum model, on the one hand we should calculate the Hubble parameter and on the other hand, we must get $q(z)$ for this fiducial model. The Hubble parameter is simply $H/H_0=1-\Omega_{m0}+\Omega_{m0} (1+z)^{3/2}$. The fiducial interaction is $q{H_0}^3=-3(1-\Omega_{m0})(1+z)H'H$. The rest of the task is to reconstruct $q(z)$ from the Hubble parameter, and then test how well the reconstructed $q(z)$ agrees with the fiducial one. If the GP method can recover both of the fiducial models and has the ability to distinguish them, we can demonstrate that GP is a valid method in the reconstruction for our propose. In the following we will create the mock data sets.

The DES is expected to obtain high quality light curves for about 4000 SNe Ia from $z=0.05$ to $z=1.2$ in the next five years. From Table $14$ in~\cite{Bernstein:2011zf} we can calculate the errors of $D$: $\sigma_{D}$ and the corresponding numbers of SNe Ia for each redshift bin. At every redshift point $z$, $D(z)_{sim}$ is sampled from the normal distribution $D(z)_{sim} \sim N(D(z)_{fid},\sigma_{D})$ where $D(z)_{sim}$ is the simulated data of $D(z)$ and $D(z)_{fid}$ is the theoretical value from the fiducial model. Thus we create the mock data sets for these two fiducial model. For each of them we reconstruct $D(z)$ and its derivatives and then apply to the reconstruction of the interaction term $q(z)$. Note that each of the reconstructed $D(z)$ and its derivatives at every redshift point is a distribution with a mean value and the error regions. However, $D(z)$ and its derivatives are not independent but correlated by a covariance between them. Using the reconstructed $D(z)$ and its derivatives also the covariance matrix between them, we can apply Monte Carlo sampling to determine the $q(z)$ in Eq.~(\ref{equa:qD}) for a given $w$ at each point $z$ which we want to reconstruct. The detailed description of the covariance matrix can also be found in~\cite{Seikel:2012uu}.

We can see from Fig.~\ref{fig:mockDLCDM} that the $D(z)$ and its derivatives are reconstructed very well from the mock data sets assuming the $\Lambda$CDM model. The dashed blue line is the mean of the reconstruction and the shaded blue regions are the $68\%$ and $95\%$ C.L. of the reconstruction. The errors for higher derivative of $D(z)$ are a litter larger than the lower ones.
As expected, at higher redshifts the errors become large due to the poor quality data in that region.

Since our reconstruction involves $D'''(z)$ whose errors are even lager in the higher redshift, and will definitely lead to somehow uncontrollable large errors in high redshift regions when apply it to the Monte Carlo sampling in Eq.~(\ref{equa:qD}). For a better show, we introduce a prefactor $(1+z)^{-n}$ to $q(z)$, that is, $\tilde{q}(z)=q(z)(1+z)^{-n}$. The choice of $n$ is somehow arbitrary, we take $n=6$ here. The motivation to introduce the prefactor is comprehensible:
we just focus on the interaction in the low and medium redshift range because the quality of observed data in the higher redshift range is so poor
that it gives a weak constraint on the reconstructed interaction. Such a pre-factor is just considered as a scale transformation with respect to redshift, which does not influence the reconstruction of $q(z)$ in the low and medium redshift ranges significantly and provides a better show in the higher redshift range. Moreover, our aim is to examine the evidence of the interaction, namely, we are testing the equality of the quantity with zero. As a consequence, we are free to do this without loss of generality like what Ref. [30] has done for the null test of the $\Lambda$CDM model. In the rest of this paper, we use $\tilde{q}$ as our interaction term to test the interaction and we mainly focus on the low and medium redshift ranges because of the poor quality of the observed data in high redshift ranges which will not give a good constraint for the interaction term.

Figure~\ref{fig:mockqLCDM} shows that the reconstructed interaction $\tilde{q}(z)$ is consistent with the fiducial $\Lambda$CDM model nicely, falling in the $1\sigma$ limit. The reconstruction in the redshift range from 0 to 0.6 is better than that in the high redshift where the error is large. Further, we see from Figs~\ref{fig:mockDdecay} and~\ref{fig:mockqdecay} that the reconstructions of the decaying vacuum model also recover the fiducial model very well, and obviously deviate from the $\Lambda$CDM model. This shows that the Gaussian processes can capture both the two models very well and correctly distinguish between them.

\subsection{Reconstruction using Union 2.1 data}

We now apply the GP method to the real data, Union 2.1 data sets~\cite{Suzuki:2011hu}, which contains 580 SNeIa data. Let us consider
\begin{equation}
m-M+5\log\left[\frac{H_0}{c}\right]-25=5\log\left[(1+z)D\right],
\end{equation}
with $H_0=70$ km/(s Mpc), following~\cite{Suzuki:2011hu}. Actually our results are not sensitive to the values of $H_0$. The values of $D$ depend on both $H_0$ and the absolute magnitude $M$. We can fix the $H_0$ and only consider the uncertainties in $M$ as adopted in Ref.~\cite{Yahya:2013xma}. We transform the distance modulus $m-M$ given in the data set to $D$ and set the theoretical initial conditions $D(z=0)=0$ and $D'(z=0)=1$. First, let us consider the decaying vacuum energy case with $w=-1$~\cite{Salvatelli:2014zta}. Figure~\ref{fig:Dunion2.1} shows the reconstructed $D(z)$ and its derivatives from Union 2.1 data. 
 The error is a little larger than that for the reconstruction of the DES survey due to the smaller number of SNe Ia and larger measurement errors. The reconstruction of the interaction $\tilde{q}(z)$ is shown in Fig.~\ref{fig:qunion2.1}. We see that both the distance $D(z)$ and interaction $\tilde{q}(z)$ are consistent with the $\Lambda$CDM model, which implies that there is no evidence for the existence of the interaction.

We now change the equation of state $w$ to see how much the differences of $w$ will influence the output of our reconstruction. We consider the $w$CDM model with $w=-1.006\pm0.045$ from the Planck 2015~\cite{Ade:2015xua}, the  CPL parametrization  $w(a)=w_0+w_a(1-a)$  with $w_0=-1.046_{-0.170}^{+0.179}$ and $w_a=0.14_{-0.76}^{+0.60}$ from HST Cluster Supernova Survey 2011~\cite{Suzuki:2011hu} and other constant equation of state: $w=-0.7$, $-0.8$, $-0.9$, $-1.1$, $-1.2$ and $-1.3$, respectively, for a comparison. All of the results are presented in Fig.~\ref{fig:qofotherw}.
Here we emphasize that we have considered the effects of the errors of $w$ on the reconstructed $\tilde{q}(z)$
in the cases of (d) and (e) in Fig.~\ref{fig:qofotherw}.
Since the constant $w$ for the $w$CDM model from Planck 2015 is very close to $-1$, so the reconstruction is almost the same as the $w=-1$ case. While for the CPL case, the error of the reconstruction is a little larger than those in other cases because of the poor constraints on $w_0$ and $w_a$. We see  that if $w$ lies between $-0.9$ and $-1.1$, $\tilde{q}(z)=0$ is captured within the $95\%$ confidence region of the reconstruction, while the interaction is shown up if the equation of state for dark energy deviates significantly from $-1$. This shows the fact that there is a degeneracy between the interaction and the equation of state of dark energy, as indicated in Eq.~(\ref{equa:qD}).

\section{ discussions and conclusions \label{sec:discussion}}

We have presented an approach to reconstruct the interaction  between dark energy and dark matter by using Gaussian processes.
To check the reliability of the GP method, we create mock data for two fiducial models, one is the $\Lambda$CDM model which
has no interaction between dark matter and dark energy, the other is a toy decaying vacuum energy model with $\rho_{DE}=3\alpha H$, which is of the interaction. It shows that the reconstruction method by using the Gaussian process works well and can capture the
features of these two models.

 We then applied the method to the real data from Union 2.1 data sets and reconstructed the distance and its derivatives, and then the
 interaction. It was found that for the decaying vacuum energy model with $w=-1$~\cite{Salvatelli:2014zta}, there is no evidence for the existence of the interaction, namely, the $\Lambda$CDM model is consistent with the Union 2.1 data sets within $1\sigma$ limits.

To check the influence of the equation of state of dark energy on the method, we have also considered several constant values of $w$  from $-0.7$ to $-1.3$.  The results show that $\tilde{q}=0$ falls in $95\%$ C.L. of the reconstruction if $w$ lies between $-0.9$ and $-1.1$. The $w$CDM model with $w=-1.006\pm0.045$ from Planck 2015 falls in this range and  $q=0$ is within the $1\sigma$ limits. This also holds for the CPL parametrization case. However, as we can see from Figure \ref{fig:qofotherw}, if $w$ deviates obviously from $-1$, the interaction exists beyond 2$\sigma$ C.L.. This reflects the degeneracy between the interaction and the equation of state of dark energy.

Note that in our reconstruction method, only the observational data on the expansion history of the universe can be used. For example, some measurements of Hubble parameter can also be combined into the above reconstruction. It is certainly of interest to develop a reconstruction method for the interaction between dark matter and dark energy, in which some kinetic data of the universe, for example, the growth factor of large scale structure, can be used for this aim.  Our paper shows that the existence of the interaction between
dark matter and dark energy found in~\cite{Salvatelli:2014zta} is mainly due to the data of redshift space distortion.




\begin{acknowledgements}
We thank Marina Seikel and Vinicius C. Busti for the helpful discussion on Gaussian processes. This work is supported by the Strategic Priority Research Program of the Chinese Academy of Sciences, Grant No.XDB09000000.
Z.K.G is supported by the National Natural Science Foundation of China Grants No.11175225 and No.11335012.
\end{acknowledgements}


\end{document}